\documentstyle[11pt,aasms4,epsfig]{article}
\begin{document}

\title{Discovery of a 6.4 keV Emission Line in a Burst from SGR 1900+14}
\author{Tod E. Strohmayer and Alaa I. Ibrahim\altaffilmark{1}}
\affil{Laboratory for High Energy Astrophysics \\ NASA's Goddard Space Flight 
Center \\ Greenbelt, MD 20771}
\altaffiltext{1}{Department of Physics, George Washington University}

\begin{abstract}		

We present evidence of a 6.4 keV emission line during a
burst from the soft gamma-ray repeater SGR 1900+14. The Rossi X-ray Timing 
Explorer (RXTE) monitored this source extensively during its outburst in the 
summer of 1998. A strong burst observed on August 29, 1998 revealed a number of
unique properties. The burst exhibits a precursor and is followed by a long 
($\sim 10^3$ s) tail modulated at the 5.16 s stellar rotation period. The 
precursor has a duration of $\approx 0.85$ s and shows both significant 
spectral evolution as well as an emission feature centered near 6.4 keV during 
the first 0.3 s of the event, when the X-ray spectrum was hardest. The 
continuum during the burst is well fit with an optically thin thermal 
bremsstrahlung (OTTB) spectrum with the temperature ranging from $\approx 40$ 
to 10 keV.  The line is strong, with an equivalent width of $\sim 400$ eV, and 
is consistent with Fe K-$\alpha$ fluorescence from relatively cool 
material. If the rest-frame energy is indeed 6.4 keV, then the lack of an 
observed redshift indicates that the source is at least $\sim 80$ 
km above the neutron star surface. We discuss the implications of the line 
detection in the context of models for SGRs.
 
\end{abstract}


\centerline{Accepted for Publication in the Astrophysical Journal Letters}


\section{Introduction}

Soft Gamma-ray Repeaters (SGRs) and Anomalous X-ray Pulsars (AXPs) are almost 
certainly young neutron stars with spin periods in the 5 - 10 s range and 
large spin down rates. 
The SGRs occasionally and unexpectedly produce short energetic ($ > 10^{41}$ 
ergs s$^{-1}$) bursts of X-ray and $\gamma$-ray radiation and two of these 
sources (SGRs 0526-66 and 1900+14) have also produced so called `giant flares',
like the famous March 5th event, with luminosities upwards of $10^{44}$ 
ergs s$^{-1}$. The persistent X-ray luminosities of both SGRs and AXPs are 
much larger than their spin down luminosities, implying some other source of 
energy powers the X-ray emission. The recent discoveries of X-ray pulsations 
from the SGR persistent X-ray counterparts (Kouveliotou et al. 1998, 1999; 
Hurley et al. 1999a) and their large period derivatives, $\dot P \approx 
10^{-10}$ s s$^{-1}$, when interpreted in the context of magnetic dipole 
radiation, have provided support for the hypothesis, first proposed 
by Duncan \& Thompson (1992), that SGRs are magnetars, neutron 
stars with dipolar magnetic fields much larger than the quantum critical field 
$B_c = m_e c^3 / e \hbar \approx 4.4 \times 10^{13}$ G (see also Paczynski
1992, Thompson \& Duncan 1995, 1996). It has also been suggested that AXPs
may be magnetars, but in a later, less active evolutionary state 
(Kouveliotou 1998).

Marsden, Rothschild \& Lingenfelter (1999) suggest that the observed variations
in $\dot P$ from SGR 1900+14 are inconsistent with magnetic dipole spin down 
and that the torque may be dominated by a relativistic wind, in which case a
dipolar field may not have to be super-critical. More recently, Harding, 
Contopoulos \& Kazanas (1999) have computed the torque due to an episodic 
particle wind and suggest that a magnetar field can still be consistent with 
the observed spin down rates and supernova remnant ages as long as the wind 
duty cycle and luminosity are within well defined limits. 

Attention has also been focused recently on accretion models for these sources
(see Marsden et al. 1999; Chatterjee, Hernquist \& Narayan 1999; and Alpar 
1999). These models suggest the spin down torques may be due to
disk accretion. However, direct evidence for such 
disks is extremely limited, and with daunting constraints on the presence of 
companions (Mereghetti 1999; and Mereghetti, Israel, \& Stella 1998; Hulleman 
2000) it seems unlikely that a binary companion could be the source of such 
material. These models also have difficulty explaining both the presence of 
bursts in SGRs and the apparently very quiet spin down of at least some of the 
AXPs (see Kaspi, Chakrabarty, \& Steinberger 1999; Baykal et al. 2000). 

Although much evidence supports the magnetar hypothesis, recent
findings have provided new challenges and it remains for 
continued observations to either vindicate or disprove the hypothesis. 
In this Letter we describe recent X-ray spectral analysis of an unusual burst 
from SGR 1900+14 which provides the first strong evidence for line emission in 
an SGR burst. Here we focus on the evidence for both spectral features and 
spectral evolution during a precursor event to the strong burst observed with 
RXTE and BATSE on August 29, 1998 UT (see Ibrahim et al. 2000). We note that 
this event was also seen by the BeppoSAX Gamma-ray Burst Monitor and Ulysses 
(Hurley et al. 2000).

\section{RXTE Observations}

On 29 August 1998, at 10:16:32.5 UT a bright burst was observed during RXTE
pointed observations of SGR 1900+14. The burst is unusual in that it showed
a rather long ($\sim 1$ s) precursor and a long decaying tail modulated with 
strong 5.16 s pulsations, similar in this respect to the giant flare which
occurred on August 27, 1998 (see Hurley et al. 1999b). Here we describe in 
detail the spectral properties of the precursor. We used data from the 
proportional counter array (PCA) in an event mode which provides the time of 
each good X-ray event to a resolution of 125 $\mu$-sec (1/8192 s) and its 
energy in one of 64 bins across the 2 - 90 keV PCA response. The event was also
observed by BATSE (Ibrahim et al. 2000). 

\begin{figure}[htb]  

\epsscale{0.8}
\plotone{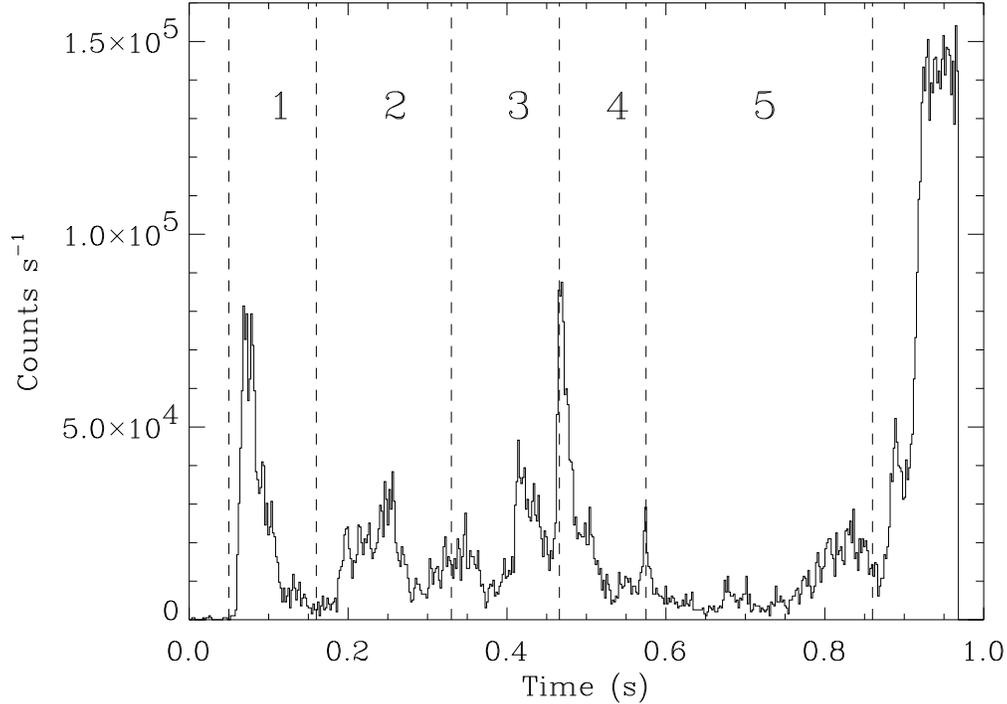}


\caption{Time history of the precursor burst to the August 29, 1998 
event from SGR 1900+14 observed with the RXTE/PCA.  The countrate is across
the full 2-90 keV PCA bandpass and the sampling rate was 512 Hz. The vertical 
dashed lines denote the intervals used in the spectral analysis. Time is 
measured from $T_0 = 51054.428171682055$ MJD (TT). The rise to $\sim 150,000$
counts s$^{-1}$ near 0.9 s represents the start of the main event.
\label{cmfig1}}
\end{figure}

Figure 1 shows the time history in the entire PCA bandpass of the precursor 
with 2 ms time resolution ($\Delta t = 1/512$ s). The precursor is rather long 
($\sim 0.85$ s) compared to typical SGR burst durations and is comprised of 
several peaks. Multi-peaked compositions are not atypical for such bursts. 
Not only is the precursor itself rather long, but the main burst itself 
lasted $\sim 3.5$ s, quite long for an SGR burst. 
In order to investigate the precursor's spectrum we accumulatd 
data in 5 independent intervals. These are labelled in Figure 1. 
We chose 5 intervals as a trade-off between having sufficient signal in each 
interval and a desire to search for spectral changes within the burst. We 
further chose the interval boundaries so that each interval would contain the 
same number of counts and therefore yield spectra of similar statistical 
quality. We used a $\sim 1000 $ s segment of preburst data as a background 
estimate. Thus our results describe the spectrum of the burst emission alone. 
We note, however, that there are very few background counts in the accumulated 
spectra. Thus, in terms of the derived spectral shape, including the presence 
of narrow spectral features, the background is essentially negligible.

We began by fitting the accumulated spectra with an optically thin thermal 
bremsstrahlung (OTTB) model including photoelectric absorption (bremss 
$\times$ wabs in XSPEC v10.0), a spectral form which has often been used
to characterize SGR bursts. We found that the OTTB model provides
an adequate description of the continuum in all the intervals. In the
first two intervals inspection of the residuals suggested a narrow excess in
the vicinity of 6.4 keV. We then added these two intervals together 
and fit the combined spectrum with the same model. To quantitatively assess 
the significance of the excess we added a narrow gaussian emission line to the 
model and evaluated the significance of the change in $\chi^2$ using the F-test
(see Bevington 1992). Since the width of the feature is of the same order as 
the PCA instrumental width we are justified in using a narrow line (ie, a 
zero-width gaussian). We also checked {\it a posteriori} that allowing a finite
width did not produce a significant decrease in $\chi^2$. Our analysis 
indicates that the narrow line significantly improves $\chi^2$ and find a 
single-trial significance for the 6.4 keV feature of $3.8 \times 10^{-5}$ using
the spectrum accumulated during intervals 1 and 2. Table 1 summarizes the 
spectral fits to all 5 intervals, both with and without the line component, 
and includes the F-test probabilities. Figure 2 
shows the residuals (Data - Model, in units of standard deviations) as a 
function of channel energy for the sum of intervals 1 and 2 using only the best
fitting OTTB continuum model (a), and the continuum plus gaussian line at 6.4 
keV (b). After modeling the feature at 6.4 keV we still see a weaker excess 
near $\approx 13$ keV. If we just model this feature while ignoring the 6.4 keV
excess, we find an F-test significance for the additional 2 parameters of only 
0.075, so statistically it is much less compelling than the 6.4 keV feature, 
however, we note that the fitted centroid is consistent with a harmonic 
relationship to the 6.4 keV feature. 

We did not find any evidence for significant excesses in the residuals from 
intervals 3 through 5. To emphasize this we show in figure 3 
the residuals from a fit to the sum of intervals 4 and 5 using the OTTB model 
which reveal no evidence for an excess. We note that the OTTB continuum is 
much softer in these intervals than the first two (see Table 1 and Ibrahim
et al. 2000).

At the PCA countrates observed during the precursor deadtime effects are not
entirely negligible. We have shown elsewhere (Ibrahim et al. 2000) that for
the observed rates during the precursor the effects of deadtime, and in this
case more crucially, pulse pile-up, are not sufficient to explain the observed
changes in the spectral continuum, nor can pile-up itself produce a narrow
spectral feature.  

The single-trial significance for the 6.4 keV feature is about $4 \times
10^{-5}$. We did analyse a number of independent intervals, so in 
estimating an overall detection significance we must pay a trials penalty. 
Although the lines were detected in the first
intervals examined we can be conservative and use 6 trials, one for each of
the five independent intervals and one for the spectrum obtained by adding
intervals 1 and 2. Even with this conservative number of trials we still 
have a significance of $2.4 \times 10^{-4}$, a robust detection.  
That the line is present during only a portion of the burst, and not in 
intervals with similar count rates only tenths of seconds apart provides a 
solid argument that the feature is not instrumental. Further, the line has 
large equivalent width (EW), much larger in fact than any previously reported 
imperfections or residuals that could plausibly be attributed to the PCA 
response matrix (FTOOLS V4.2 and PCA response matrix generator MARFRMF V3.2.1; 
Jahoda et al. 2000).  All these arguments provide very strong evidence that 
the line feature at 6.4 keV is real and therefore intrinsic to the source. 

\begin{figure}[htb]  

\epsscale{0.8}
\plottwo{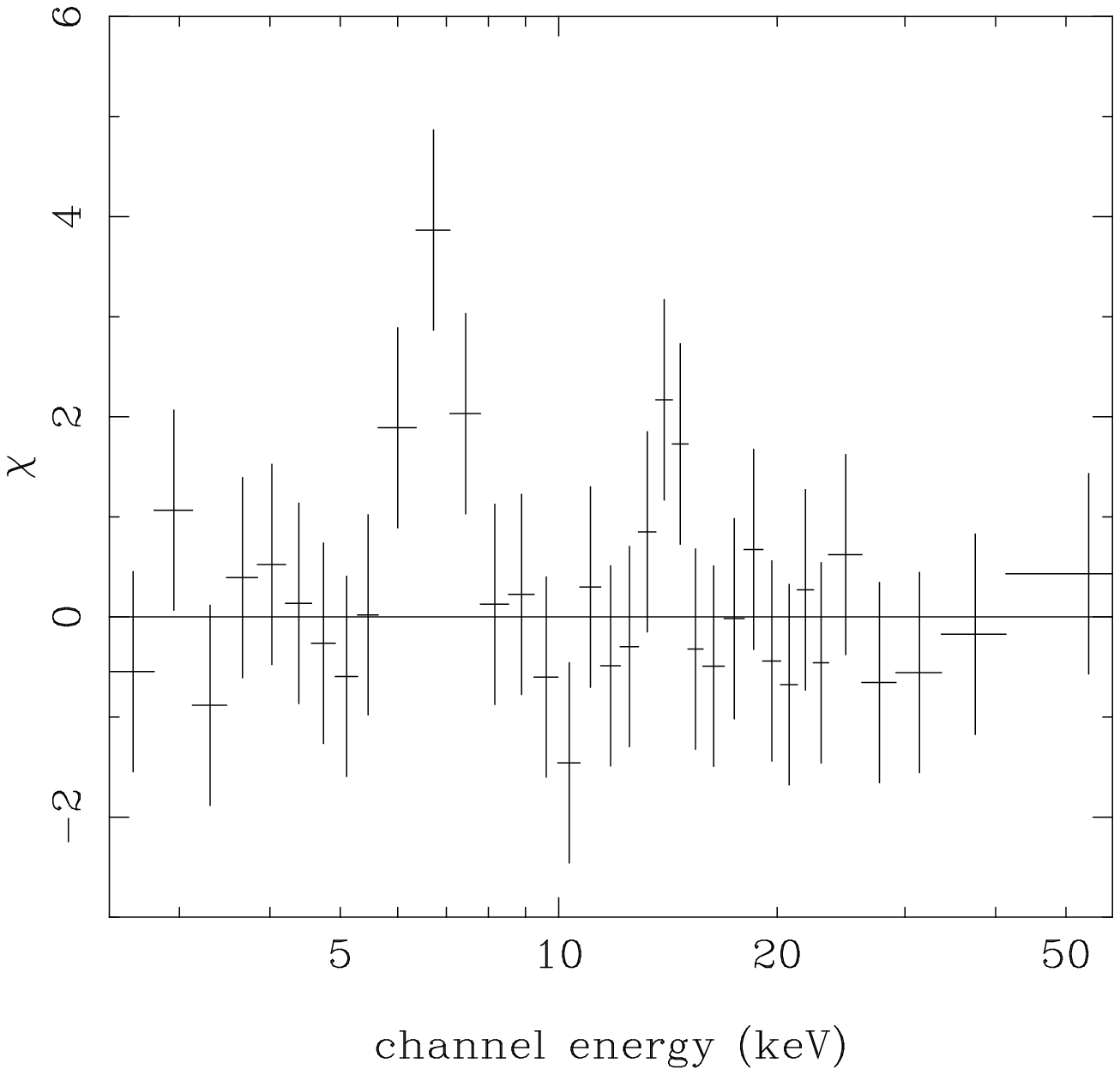}{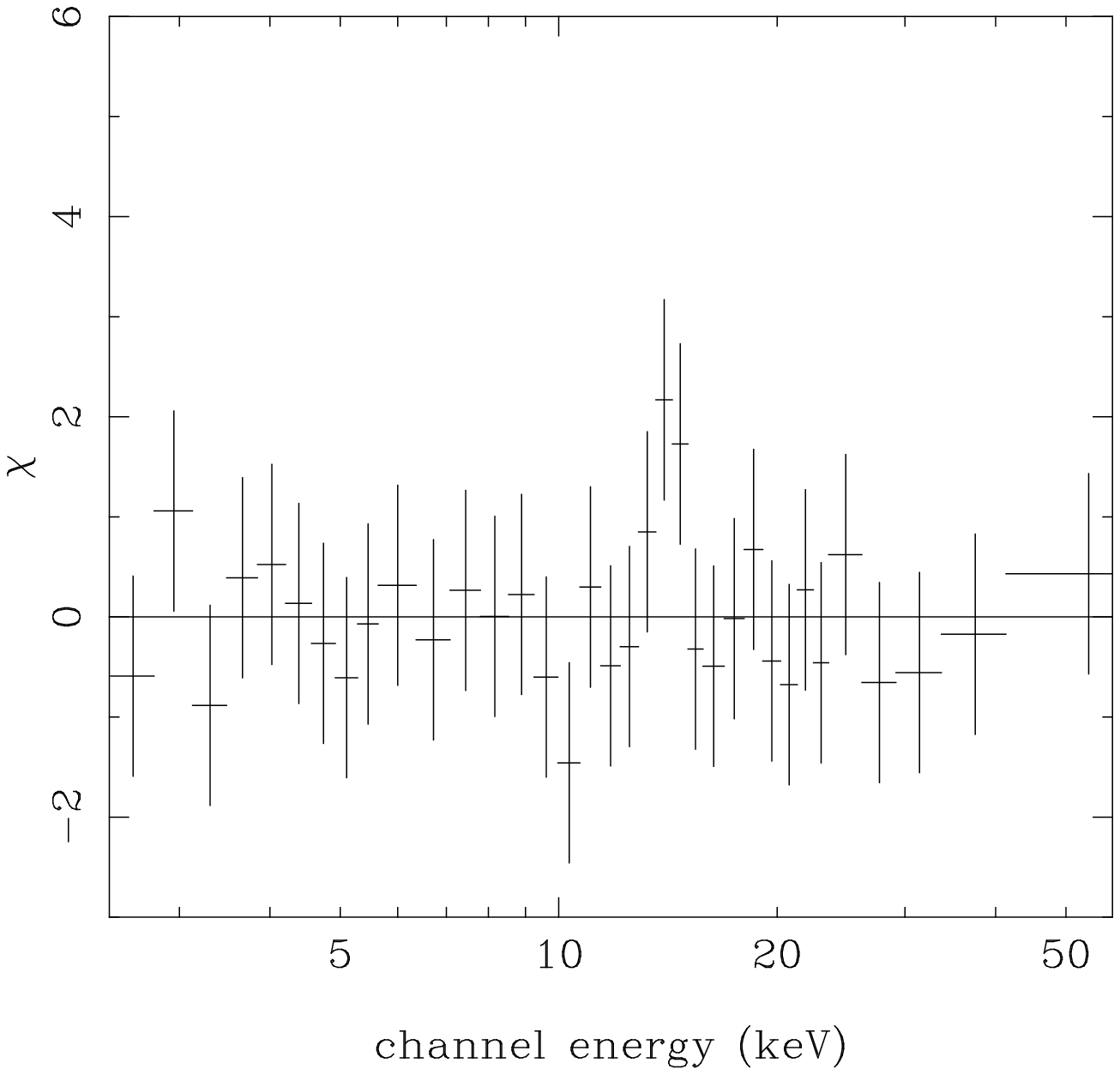}
\caption{Residual plots (Data - Model, in units of standard 
deviations) using the OTTB fit from the sum of the first two intervals in the 
precursor. The left panel shows the residuals using only the best fitting 
continuum model and shows a strong excess at $\sim 6.4$ keV, while the right
panel includes the continuum and the gaussian line at 6.4 keV.
\label{cmfig2ab}}
\end{figure}

\section{Summary and Discussion}

The spectral behavior reported above is unique
in several ways. First, a narrow spectral feature in an 
SGR burst has not to our knowledge been previously reported, and second, the 
burst also shows a dramatic spectral softening. Strohmayer \& Ibrahim (1998) 
demonstrated that some bursts from SGR 1806-20 also show significant
spectral evolution. Interestingly, the earlier result from SGR 1806-20 is 
similar in that the trend is for hard to soft evolution. Finally, the line 
feature is only present during the hardest spectral intervals, suggesting a 
possible connection between these properties. 

The presence of line emission during an SGR burst raises a host of interesting 
questions concerning the production mechanism and has many implications 
for models of SGRs.  It is beyond the scope of this paper to attempt an 
exhaustive description of possible models, rather, we qualitatively discuss 
several possibilities. 

Of the various models proposed for SGR bursts the most often discussed have 
been the magnetar model (Thompson \& Duncan 1995) and models based on sudden
accretion (see Colgate \& Petschek 1981; Epstein 1985; Colgate \& Leonard 1993;
Katz, Toole \& Uhruh 1994). To date there has been little direct evidence to
support accretion scenarios. An intriguing possibility,
however, is that the line is due to fluorescence of 
iron in relatively cool material located near the neutron star. Such features 
have been observed in a number of astrophysical systems, including accreting 
X-ray pulsars and magnetic CVs (see White, Nagase \& Parmar 1993; Ezuka \& 
Ishida 1999; and references therein). 
In the X-ray pulsars some portion of the line EW is correlated with the
observed absorbing column, $n_H$, indicating that some of the line
EW is due to fluorescence in the absorbing material, most likely the stellar 
wind from the secondary. In some circumstances there is evidence for an 
uncorrelated EW indicating an additional source of fluorescing material and it 
has been suggested that matter might accumulate in the Alfven shell (Inoue 
1985; Basko 1980) or perhaps an accretion disk (see Bai 1980). 

The appearance of lines in the disk accretors, 
Her X-1 and Cen X-3 indicates that a disk can also be an efficient reprocessor.
The fluorescent lines from accreting X-ray 
pulsars are strong with EWs greater than 100 eV. This is at least 
qualitatively similar to the feature described here. 
If the line is due to 6.4 keV iron emission in the rest frame then the 
fluorescing material cannot be the stellar surface itself because of the 
absence of any significant redshift. 
Assuming the rest-frame energy of the line is 6.4 keV we can place a 90 \% 
confidence lower limit of $\approx 80$ km on the altitude of the fluorescing 
material above a typical neutron star ($M = 1.4 M_{sun}$ and $R = 10$ km). 
Note that this estimate also ignores any possible
effects of a magnetar strength field on the line spectrum. However, the 
question remains as to the source of the fluorescing material.

\begin{figure}[htb]  

\epsscale{0.6}
\plotone{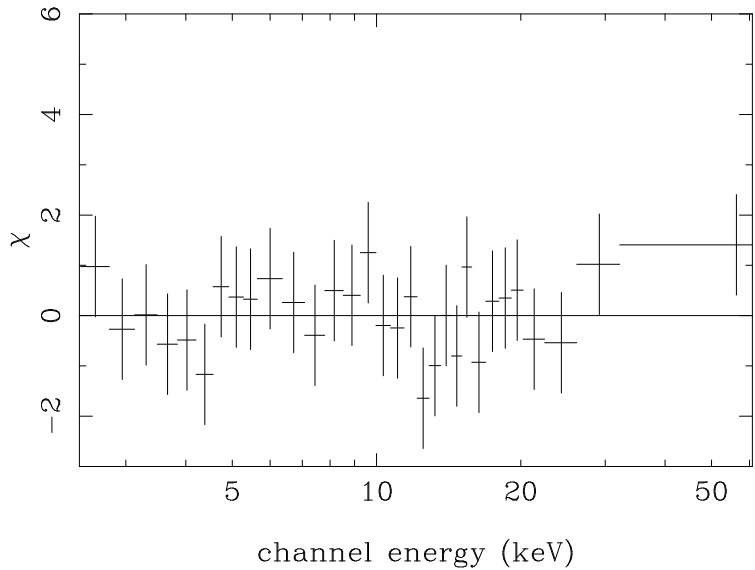}
\caption{Residual plot (Data - Model, in units of standard 
deviations) using the OTTB fit from the sum of intervals 4 and 5 of the 
precursor. Their is no evidence for an excess at 6.4 keV in these intervals.
\label{cmfig2c}}
\end{figure}

SGR giant flares likely produce a hot, optically thick electron-positron 
plasma in the magnetosphere (Thompson \& Duncan 1995). 
Energy from this radiating plasma will likely 
be conducted into the surface of the star, which increases the crust 
temperature, and may also blow off a baryon wind.
(Thompson \& Duncan 1995). It is conceivable that an ablation process 
caused either by the precursor itself, or perhaps by the giant
flare which occurred on August 27th, 1998 may have ejected
enough iron-rich material into the neutron star environment to produce the 
flourescence. Somewhat related processes have been discussed in the context of 
the putative iron lines from gamma-ray burst afterglows 
(see Piro et al. 1999; Ghisellini, Lazzati, \& Campana 1999). The SGR 
phenomenon reported here could conceivably be similar but on a smaller scale.

The apparent changes in spin down rate observed in SGR 1900+14 (Marsden, 
Rothschild \& Lingenfelter 1999) have been used to argue for the presence of
circumstellar material to produce the spin down torques. This material 
might be co-moving ejecta from the supernova explosion (Marsden et al. 1999),
or perhaps a fall back accretion disk (see Chatterjee et al. 1999; Alpar 1999).
It is possible that such material could be the source of a fluorescence line.
Another clue would seem to be the dissapearance of the line as the continuum
softens. This appears to be qualitatively consistent with fluorescence as the
line mechanism, as the line strength will depend on the number of photons above
the Fe K-edge absorbed in the fluorescing material, and as the spectrum hardens
(softens) this quantity will increase (decrease).

Although the weaker excess at $\sim 13$ keV is not compellingly significant 
on its own, the harmonic relationship with the 6.4 keV feature is suggestive 
of cyclotron emission from accreting pulsars (see Nagase 1989; Dal Fiume et al.
1999). For magnetar-strength fields {\it electron} cyclotron transitions would 
lie above an MeV, well above the bandpass of RXTE/PCA.  However, Duncan
\& Thompson (2000) have recently pointed out the following mechanism 
involving {\it proton} and {\it alpha particle} cyclotron transitions.
In the magnetar model, the giant flare on August 27th produces a hot 
($T\sim$ MeV) trapped fireball in the stellar magnetosphere. Heavy nuclei 
are ejected from the stellar surface into this fireball, where they 
photodissociate and subsequently settle to form a thin layer of light 
elements on the stellar surface.  In the August 29 precursor which follows, 
radiative heating of the star's surface gives rise to emission at the 
cyclotron fundamental of each ion, separated by a factor two in frequency.  
If the 6.4 keV line is the He$^4$ cyclotron fundamental, then the implied 
surface field strength is $2.6\times 10^{15}$ G.  This estimate takes into 
account a redshift correction from the surface of a canonical neutron star.  
The alternative interpretation of the lines as first and second proton 
harmonics in $1.3 \times 10^{15}$ G seems less plausible.  In any case, 
the surface field strength somewhat exceeds the dipole field strength deduced 
from the spindown of SGR 1900+14; but note that the multi-peaked pulse profile 
in the tail of the Aug. 27th flare gives evidence for stronger higher-order 
multipoles (Thompson et al. 2000).  More work is required to determine if 
this model can account for the observed line EW and other details, but it 
is an intriguing possibility.

It is sometimes difficult to draw firm conclusions based on
a single example, however, our results argue strongly for detailed spectral 
studies of the whole sample of bursts observed with RXTE from SGR 1900+14, a 
project we are pursuing. Studies with instruments possessing greater 
spectral resolution (such as Chandra and XMM, if they can handle the 
high fluxes) may hold great promise for further testing the magnetar hypothesis
for SGRs.

\acknowledgements

A. I. is grateful to Samar Safi-Harb and Eamon Harper for many useful 
discussions. We also thank Jean Swank, Craig Markwardt, Dave Marsden, Robert 
Duncan and Chris Thompson for their input and comments. 

\vfill\eject

\vfill\eject

\vskip 10pt
\hoffset -0.75in

\begin{table}

\caption{Spectral results for the August 29, 1998 precursor}
\begin{tabular}{ccccccccc}
\tableline
Interval & kT (keV) & $n_H$ ($10^{22}$ cm$^{-2}$) & $E_c$ (keV) & f (cm$^{-2}$ 
s$^{-1}$) & EW (eV) & $P_{F-test}$ & $\chi^2$ & dof \cr
\tableline

1+2 (line) & 33.8 $\pm$ 4.1 & 3.8 $\pm 0.7$  & 6.48 $\pm 0.14$ & 0.24 
$\pm 0.05$
& 414 $\pm 95$ & 3.8 $\times 10^{-5}$ & 18.5 & 30 \cr
1+2 (no line) & 26.9 $\pm$ 2.6 & 4.9 $\pm 0.7$  & & & & & 36.5 & 32 \cr
1 (line) & 42.6 $\pm$ 9.2 & 3.9 $\pm 1.0$  & 6.40 $\pm 0.22$ & 0.30 $\pm 0.11$ 
& 398 $\pm 140$ & 1.68 $\times 10^{-2}$ & 19.5 & 26 \cr
1 (no line) & 34.8 $\pm$ 5.9 & 5.0 $\pm 1.1$  & & & & & 26.7 & 28 \cr
2 (line) & 23.8 $\pm$ 3.1 & 3.3 $\pm 1.3$  & 6.56 $\pm 0.18$ & 0.20 $\pm 0.06$ 
& 440 $\pm 130$ & 1.1 $\times 10^{-3}$ & 13.5 & 23 \cr
2 (no line) & 18.9 $\pm$ 2.4 & 4.9 $\pm 1.2$  & & & & & 23.3 & 25 \cr
3 & 15.9 $\pm$ 1.4 & 4.1 $\pm 0.8$  & & & & & 34.5 & 32 \cr
4 & 11.8 $\pm$ 0.9 & 3.9 $\pm 0.8$  & & & & & 16.7 & 27 \cr
5 & 20.4 $\pm$ 2.1 & 5.1 $\pm 1.0$  & & & & & 29.0 & 29 \cr
\tableline
\end{tabular}
\vspace{0.5 cm}
\hoffset -0.75in

\end{table}
\vfill\eject
\hoffset -0.750in

\end{document}